\documentclass[11pt]{article}
\usepackage{amsmath,amsfonts,amssymb}
\usepackage{graphicx,color}
\usepackage{bm}
\usepackage{multirow}
\usepackage{cite}
\usepackage{color}
\usepackage[bookmarks, breaklinks, colorlinks,urlcolor=black, citecolor=red, linkcolor=blue]{hyperref}

\usepackage{graphics}
\usepackage{mathrsfs}
\usepackage{slashed}
 \usepackage[normalem]{ulem}

  \long\def\hlpbp#1///{\color{magenta} #1 \color{black}}
 \definecolor{darkgreen}{cmyk}{1,0,1,0.4}

\def\com2#1{\textcolor{red}{\it{#1}}}
\long\def\rpl#1!!#2!!{\color[cmyk]{.3,.5,.9,.1}
  \ensuremath{\rhd} #1 \ensuremath{\lhd}
  \color{blue} #2 \color{black}}

\def\3Eqs#1#2#3{Eqs.\ (\ref{#1}), (\ref{#2}) and (\ref{#3})}

\title{\bf Enhancement of the $\bm{W}$ boson mass in the Georgi-Machacek model}

\author{ 
{\bf Poulami Mondal}\thanks{E-mail: poulami.mondal1994@gmail.com} \\ 
Department of Physics, University of Calcutta, \\
92 Acharya Prafulla Chandra Road, Kolkata 700009, India}

\date{April 2022}

\begin{document}

\maketitle

\renewcommand*{\thefootnote}{\fnsymbol{footnote}}

%\vspace{5mm}

\begin{abstract}

The recent CDF-II measurement of the mass of the $W$ boson shows a significant tension with the Standard Model
(SM) expectation, even when averaged with earlier measurements. The tension can be explained 
in the framework of any beyond-SM dynamics that contributes substantially to the oblique
$S$ and $T$ parameters. As a typical example of such beyond-SM physics, we try to explain the tension in the framework of the Georgi-Machacek model that keeps 
$\rho=1$ at the tree-level, and explore the parameter space that can accommodate the $W$ mass shift. We find that a sizeable split between the custodial triplet and 5-plet of scalars can help alleviate the tension
to a large extent, and raise the $W$-mass by about 30 MeV over its SM value.
\end{abstract}

%%%%%%%%%%%%%%%%%%%%%%%%%%%%%%%%%%%%%%%%%%%%%%%%%%%%%%%%%%%%%%%%%%%%%%%%%%%%%%

\setcounter{footnote}{0}
\renewcommand*{\thefootnote}{\arabic{footnote}}

%%%%%%%%%%%%%%%%%%%%%%%%%%%%%%

%%%%%%%%%%%%%%%%%%%%%%%%%%%%%%
\section{Introduction}  
%%%%%%%%%%%%%%%%%%%%%%%%%%%%%%
Recently, the CDF Collaboration published a new measurement of the $W$-boson mass, $M_W$, at the 
Fermilab Tevatron collider \cite{CDF:2022hxs} with a sample of about 4 million $W$-bosons:
\begin{equation}
\left. M_W \right\vert_{\text{CDF}}= 80434 \pm 9 ~ {\rm MeV} \,,
\label{CDF}
\end{equation}
where the systematic and the statistical errors have been added in usual quadrature. The number is an average of
6 different measurements, but the interesting point is that the central value of the most precise measurement,
the one coming from the transverse mass of the $W$ in the muonic final state, is way above
the average, while all the other 5 are below the average. Entertaining the theoretical possibility that the former may be 
an outlier, or the errors have been underestimated, we also perform a sort of naive averaging over the other 5 data points, 
and find
\begin{equation}
\left. M_W \right\vert_{\text{CDF w/o $m_T(\mu)$}}= 80429 \pm 10 ~ {\rm MeV} \,,
\label{CDFwomT} 
\end{equation}
so that the tension is still very significant. 

The numbers may be compared with 
the other measurements from LEP, D\O-II, LHCb, and ATLAS~\cite{ALEPH:2005ab,D0:2012kms,LHCb:2021bjt}:
\begin{subequations}
\label{other-expt}
\begin{eqnarray}
\label{lep} \left. M_W \right\vert_{\text{LEP-2}} &=& 80376 \pm 33 ~ {\rm MeV}\,,\\
\label{dzero} \left. M_W \right\vert_{\text{D\O-II}} &=& 80375 \pm 23 ~ {\rm MeV} \,,\\
\label{lhcb} \left. M_W \right\vert_{\text{LHCb}} &=& 80354 \pm 32 ~ {\rm MeV} \,, \\
\label{atlas} \left. M_W \right\vert_{\text{ATLAS}} &=& 80370 \pm 19 ~ {\rm MeV} \,.
\end{eqnarray}
\end{subequations}
The averages, with and without the CDF-II measurement, are given by
\begin{subequations}
\label{average}
\begin{eqnarray}
\label{nocdf} \left. M_W \right\vert_{\text{Avg.\ w/o CDF}} &=& 80370 \pm 12 ~{\rm MeV} \,,\\
\label{cdfmtmu} \left. M_W \right\vert_{\text{Avg.\ with CDF, with $m_T(\mu)$}} &=& 80411 \pm 8 ~{\rm  MeV} \,, \\
\label{cdfnomtmu} \left. M_W \right\vert_{\text{Avg.\ with CDF, w/o $m_T(\mu)$}} &=& 80402 \pm 8 ~{\rm  MeV} \,,
\end{eqnarray}
\end{subequations}
where we have assumed zero correlation among the measurements, so that the combined standard deviation 
is given by
\begin{equation}
\frac{1}{\sigma^2} = \sum_i \frac{1}{\sigma_i^2}\, \,,
\end{equation}
and the central value is just the weighted mean of the individual numbers:
\begin{equation}
\bar{x} = \frac{ \sum_i x_i/\sigma_i^2}{\sum_i 1/\sigma_i^2}\,.
\end{equation}

This measurement shows a significant deviation, of more than $6\sigma$, from the Standard Model (SM) prediction
 \cite{ParticleDataGroup:2020ssz}:
\begin{equation}
\left. M_W \right\vert_{\text{SM}}= 80361 \pm 7 ~{\rm MeV} \,.
\label{SM}
\end{equation}

This has led to a burst of attempts in the recent literature to explain the tension, either in the framework of 
SM effective theories, or in the framework of specific models.
The SM electroweak fit is performed in terms of the four input parameters, namely, the fine structure constant $\alpha$, 
the Fermi coupling $G_F$, the mass of the $Z$-boson $M_Z$, and the Higgs boson mass $M_h$. The tree-level
prediction for $M_W$ is modified by the radiative corrections, as shown in Eq.\ (\ref{SM}). With new degrees of 
freedom, there are more corrections to the $W$ propagator, which can be parametrised by the oblique parameters
$S$, $T$, and $U$ \cite{Peskin:1991sw,Davoudiasl:2015bua,Altarelli:1994iz,Boughezal:2005eb}:
\begin{equation}
\Delta M_W = \frac{\alpha M_W}{2(c_W^2-s_W^2)}\, \left (-\frac{1}{2}S + c_W^2 T + \frac{c_W^2 - s_W^2}{4s_W^2}U
\right)\,.
\label{mWST}
\end{equation}
Models with an extended scalar sector are favoured in the sense that the self-energy corrections are only logarithmic 
in scalar masses, and hence they are less tightly constrained from experimental data. For example, we may refer to 
Refs.\ \cite{Fan:2022dck} and \cite{Du:2022brr}, where the former discusses the consequences of two-Higgs doublet
models on $\Delta M_W$, and the latter focusses on an extension of the Georgi-Machacek model with 
right-handed neutrinos.

Of course, the SM itself contributes to nonzero values of the oblique parameters at the one-loop level. 
For example, fermions break custodial symmetry and a positive contribution to the $T$ parameter comes from
$M_{\rm top} \gg M_{\rm bottom}$. We set the scale so as the oblique
parameters are zero within the SM (with $M_{\rm top}\approx 173$ GeV and $M_h\approx 125$ GeV), 
Eq.\ (\ref{mWST}) constrains the new physics effects in $S$, $T$, and $U$. One may set $U\approx 0$ if the 
new degrees of freedom are much heavier compared to $M_Z$. This need not be the case always, 
but in this work, we will not venture into the $U\neq 0$ parameter space.

In models that only extend the scalar sector of the SM, $S$ can hardly be too large, as the contribution of scalars 
to $S$ is logarithmic in nature. Furthermore, any model that preserves custodial 
symmetry\footnote{For details about various types of custodial symmetries, see Ref.\ \cite{Kundu:2021pcg}.} 
naturally has $T=0$, so the parameter space becomes much more tractable in terms of $S$ only.
Models with SU(2) singlet and doublet scalars automatically keep $T=0$. However, there exists another class 
of models, where the custodial symmetry is forced by the mutual cancellation of custodial symmetry breaking effects 
coming from more than one such multiplets.
One such example is the Georgi-Machacek (GM) model \cite{Georgi:1985nv},  where one complex and 
one real triplet of scalars get identical vacuum expectation values and hence conspire to keep $T=0$. The tractability 
of the parameter space motivates us to study the effect of the GM model on $\Delta M_W$.

In this paper, we will see how far the $M_W$-tension can be alleviated in the framework of the conventional GM
model, which will be very briefly described 
in Section \ref{sec:GM}. Our results are shown in Section \ref{sec:results}, and Section 
\ref{sec:conclude} concludes the paper.

%%%%%%%%%%%%%%%%%%%%%
\section{The Georgi-Machacek model} \label{sec:GM}
%%%%%%%%%%%%%%%%%%%%%
The GM model \cite{Georgi:1985nv} extends the scalar sector of the SM by adding one 
real $SU(2)_L$ triplet $\xi$ and one complex $SU(2)_L$ triplet $\chi$ with hypercharges $Y = 0$ and 
$Y = 2$ respectively. The scalar fields of this model can be represented in the forms of a bi-doublet 
$\Phi$ and a bi-triplet $X$ as

\begin{eqnarray}
\Phi =
\begin{pmatrix}
\phi^{0*} & \phi^+\\
-\phi^{+*} & \phi^0
\end{pmatrix} ,\hspace{0.5 cm}
X=
\begin{pmatrix}
\chi^{0*} & \xi^+ & \chi^{++} \\
-\chi^{+*} & \xi^0 & \chi^+ \\
\chi^{++*} & -\xi^{+*} & \chi^0
\end{pmatrix} \,.\hspace{0.5 cm}
\label{fieldsGM1}
\end{eqnarray}

The most general gauge-invariant scalar potential with these fields that conserves the custodial SU(2) 
symmetry is given by \cite{{Keeshan:2018ypw}}:
\begin{eqnarray}
\nonumber
V(\Phi,X) &=& \frac12\, {\mu_2^2}\, {\rm Tr} (\Phi^\dagger \Phi) + \frac12\, {\mu_3^2}\, {\rm Tr} (X^\dagger X) +
 \lambda_1\, \left[ {\rm Tr} (\Phi^\dagger \Phi)\right]^2 + \lambda_2\, {\rm Tr} (\Phi^\dagger \Phi)\, 
 {\rm Tr} (X^\dagger X) + \lambda_3\, {\rm Tr} (X^\dagger X X^\dagger X)
\\
\nonumber
& &  +\lambda_4\, \left[ {\rm Tr} (X^\dagger X)\right]^2 - \lambda_5\, {\rm Tr}(\Phi^\dagger \tau^a \Phi \tau^b)\, 
{\rm Tr} (X^\dagger t^a X t^b) - M_1\, {\rm Tr}(\Phi^\dagger \tau^a \Phi \tau^b)\, (U X U^\dagger)_{ab}
\\
& &  -M_2\, {\rm Tr}(X^\dagger t^a X t^b)\, (U X U^\dagger)_{ab} \,,
\label{GMpot1}
\end{eqnarray}
where $\tau^a = \frac12 \sigma^a$ ($\sigma_a$ are the Pauli matrices) and $t^a$ are the $2\times 2$ 
and $3\times 3$ SU(2) generators respectively.

The vacuum expectation values are set as follows, where the superscripts $r$ and $i$ denote the real and the 
imaginary parts of the fields respectively:
\begin{eqnarray}
\phi^0 \rightarrow \frac{v_\phi}{\sqrt{2}}+\frac{\phi^{0,r}+i \phi^{0,i}}{\sqrt{2}}, \hspace{0.5 cm} \chi^0 \rightarrow v_\chi+\frac{\chi^{0,r}+i \chi^{0,i}}{\sqrt{2}}, \hspace{0.5 cm} \xi^0 \rightarrow v_\xi+\xi^{0,r}\,,
\label{vevsGM1}
\end{eqnarray}
with
\begin{equation}
v_\chi = v_\xi\,.
\label{forceCS}
\end{equation}

The physical fields form a fiveplet, a triplet, and two singlets according to their transformations under the
custodial SU(2) symmetry. There are one doubly charged, two singly charged (plus their antiparticles), 
one CP-odd neutral, and three CP-even neutral scalars. 
The masses within each custodial multiplet are degenerate (this is a feature of the minimal GM model,
the extended GM models \cite{Kundu:2021pcg} do not respect this degeneracy). 
We denote them by $m_5$ and $m_3$ for the fiveplet and the triplet respectively.

One may note that Eq.\ (\ref{forceCS}) is what makes the GM model attractive. If one introduces a splitting
between $v_\chi$ and $v_\xi$, $T$ becomes nonzero but the GM model loses its most attractive feature. 
This relationship holds even under higher-order radiative corrections for a pure SU(2) gauge theory.
With a gauged hypercharge U(1), Eq.\ (\ref{forceCS}) no longer holds; rather, it acquires a divergent contribution 
at one-loop \cite{{Gunion:1990dt}}. That is why we do not entertain the possibility of a tree-level
splitting between $v_\chi$ and $v_\xi$, and we do not introduce any other field except those required for the 
conventional and minimal GM model.

The mass matrix for the CP-even states can be diagonalised by the mixing angle $\vartheta$ as
\begin{equation}
h = \cos\vartheta \, \phi^{0,r} - \sin\vartheta \, H_1^{0\prime}, \qquad 
H = \sin\vartheta \, \phi^{0,r} + \cos\vartheta \,H_1^{0\prime},
\label{alpha}
\end{equation}
where
\begin{equation}
H_1^{0\prime}=\sqrt{\frac{1}{3}}\xi^{0,r} + \sqrt{\frac{2}{3}}\chi^{0,r}\,.
\label{alpha2}
\end{equation}

%%%%%%%%%%%%%%%%%%%%%%%%%%%%%%%%%%%%
\section{New physics contributions to ${\Delta M_W}$} \label{sec:results}
%%%%%%%%%%%%%%%%%%%%%%%%%%%%%%%%%%%%%
The shift in the $W$ mass, $\Delta M_W$, is defined to be the difference between the experimental average
and the SM expectation, with the errors/uncertainties added in quadrature. Thus,
\begin{equation}
\Delta M_W = \left. M_W\right\vert_{\text{Average}} - \left. M_W\right\vert_{\text{SM}} = 50 \pm 11 \, MeV \,,
\label{deltamWCDFmT}
\end{equation}
where we have used the average as shown in Eq.\ (\ref{cdfmtmu}). This goes down to $41\pm 11$ MeV if we 
neglect the $m_T(\mu)$ number from CDF-II. We stress that one should take care of all the data and 
use an average, rather than only the CDF-II number, as has been done in some works in the literature.

In the GM model, $S$ is the only relevant oblique parameter, and may be written in terms of the $Z$ boson and 
the photon self-energies as \cite{Peskin:1991sw}
\begin{eqnarray}
S &=& \frac{4s_W^2 c_W^2}{\alpha M_Z^2} \, \left[ 
\Pi_{ZZ}(M_Z^2)-\Pi_{ZZ}(0) - \frac{c_W^2-s_W^2}{s_W c_W} \Pi_{Z \gamma}(M_Z^2)-\Pi_{\gamma \gamma}(M_Z^2)
\right] \nonumber \\
& \simeq & \frac{4s_W^2 c_W^2} {\alpha} \, \left[
\Pi^\prime_{ZZ}(0)-\frac{c_W^2-s_W^2}{s_W c_W} \Pi^\prime_{Z \gamma}(0)-\Pi^\prime_{\gamma \gamma}(0)
\right] \,,
\label{S}
\end{eqnarray}
where $s_W\equiv \sin\theta_W$, $c_W\equiv\cos\theta_W$, $\theta_W$ being the Weinberg angle, 
and $\Pi^\prime = d\Pi(p^2)/dp^2$.
Setting $U=T=0$, $\Delta M_W$ can be calculated directly from Eq.\ (\ref{mWST}), 
 and putting in all the numerical values, one gets
\begin{equation}
\Delta M_W~{\rm (MeV)} \approx -265\, S\,.
\end{equation}
The analytical expression for $S$ is rather cumbersome, and is given in Ref.\ \cite{Keeshan:2018ypw}.
It turns out that $S$ depends on four parameters, namely, $m_3$ and $m_5$, the masses of the custodial triplet and
the 5-plet scalars respectively, the mixing angle $\vartheta$, and $v_\chi$.

\begin{figure}[htbp]
\includegraphics[width=6cm, angle=270]{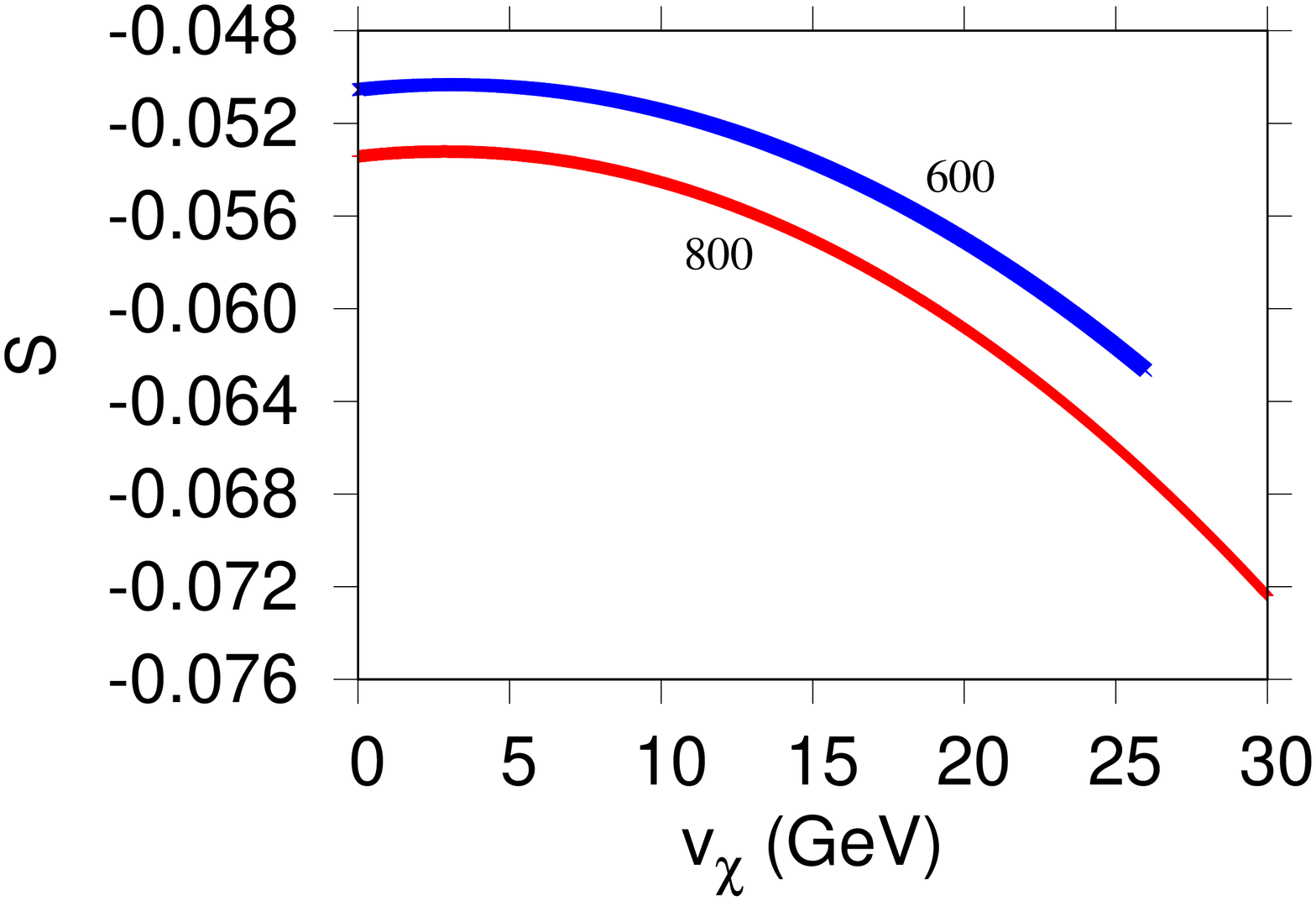} 
\includegraphics[width=6cm, angle=270]{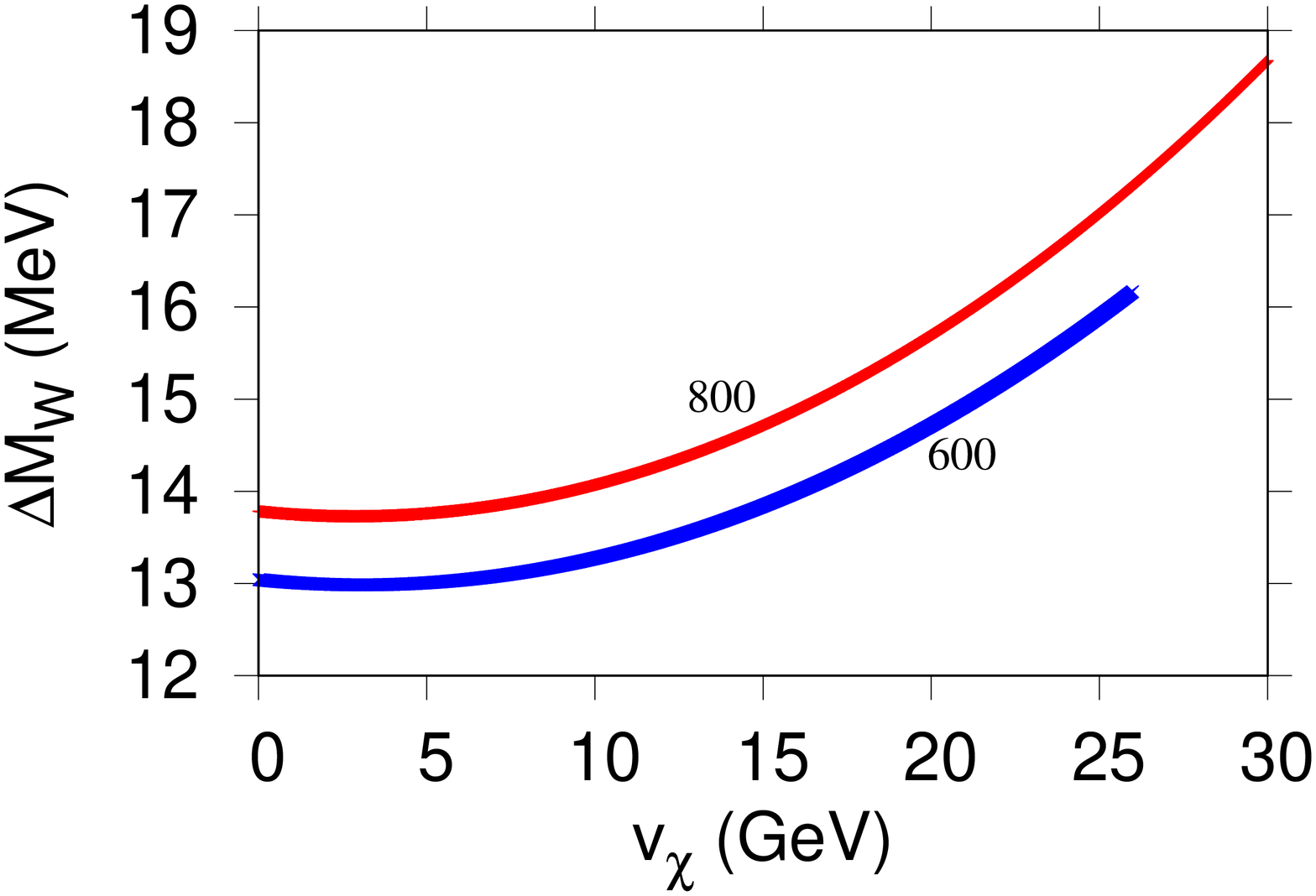} 
\caption{\small{$S$ (left) and $\Delta M_W$ (right) as a function of the triplet vacuum expectation value
$v_\chi$, for different values of $m_3 (=m_5)$. The blue and red lines are for $m_3=m_5=600$ and 800 GeV,
respectively. The mixing angle $\vartheta$ as defined in Eq.\ (\ref{alpha}) has been kept fixed at 
$\sin\vartheta = 0.31$.}}
\label{plot} 
\end{figure}

First, let us discuss a simplified version of the GM model to have a qualitative idea. 
In this version, we take all new scalars belonging to the custodial triplet and 5-plet
to be degenerate, so that $m_3=m_5$.
We show, in Fig.\ \ref{plot},  how $S$ and $\Delta M_W$ vary with the triplet vacuum expectation value $v_\chi$ in a 
simplified version of the GM model. The model parameters have been chosen in such a way that the lightest 
CP-even neutral scalar $h$ is at 125 GeV. The mixing 
angle $\vartheta$, as defined in Eq.\ (\ref{alpha}), 
is fixed at $\sin\vartheta = 0.31$, so that $h$ is dominantly a doublet and satisfies all the LHC 
constraints. 
For the other custodial singlet, $H$, we take $m_H = 150$ GeV.\footnote{Such a choice is motivated 
by some hints of a spin-0 resonance that decays to $\gamma\gamma$, see, {\em e.g.}, Ref.\ \cite{Richard}.}

At this point, one must take into account the direct and indirect constraints on the parameter space of the GM model. 
The indirect constraints come mostly from the flavour observables, through the radiative effects mediated by the 
charged Higgs belonging to the custodial triplet. These effects increase with increasing $v_\chi$, which controls 
the SU(2)$_L$ doublet component of the charged Higgs. 
Thus, observables like $R_b=\Gamma(Z\to b\bar{b})/
\Gamma(Z\to {\rm hadrons})$, branching ratios for $b\to s\gamma$ and $B_s\to\mu^+\mu^-$, and $B_s$--$
\overline{B}_s$ mass difference, contribute towards constraining the parameter space \cite{Hartling:2014aga}. 
The direct, and often the tighter, constraint on $v_\chi$ comes from the non-observability of the decay
 $H^{++}\to W^+W^+$ \cite{CMS:2017fhs}, which restricts $v_\chi$ to be less than 
$30 (25)$ for $m_5=800 (600)$ GeV. 

 With all the constraints imposed, we see from Fig.\ \ref{plot} that the maximum enhancement of $M_W$ is of the order
 of 20 MeV, and thus such a simple model is not enough to have a reasonable agreement with the data.
 
\begin{figure}[htbp]
\includegraphics[width=6cm, angle=270]{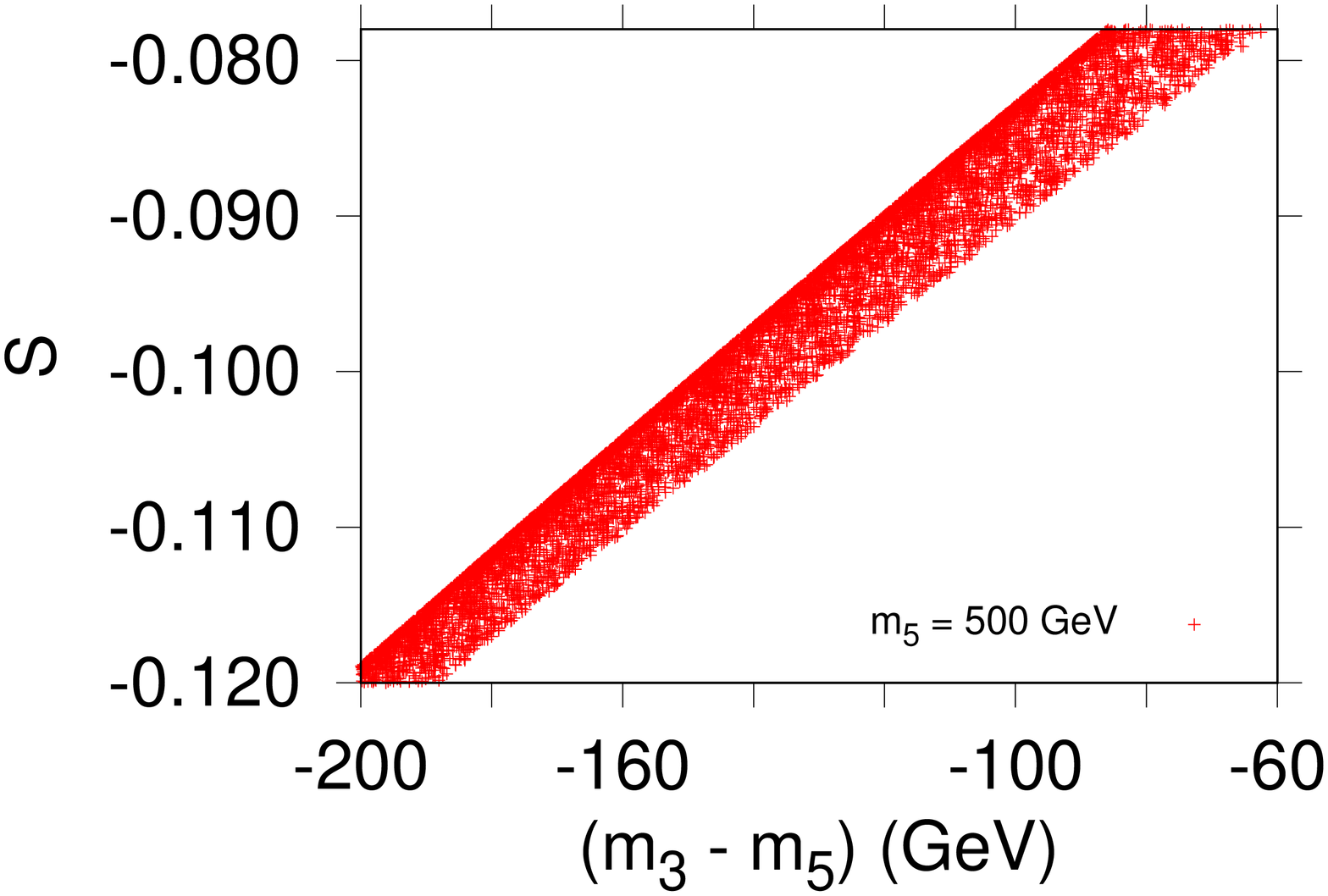} 
\includegraphics[width=6cm, angle=270]{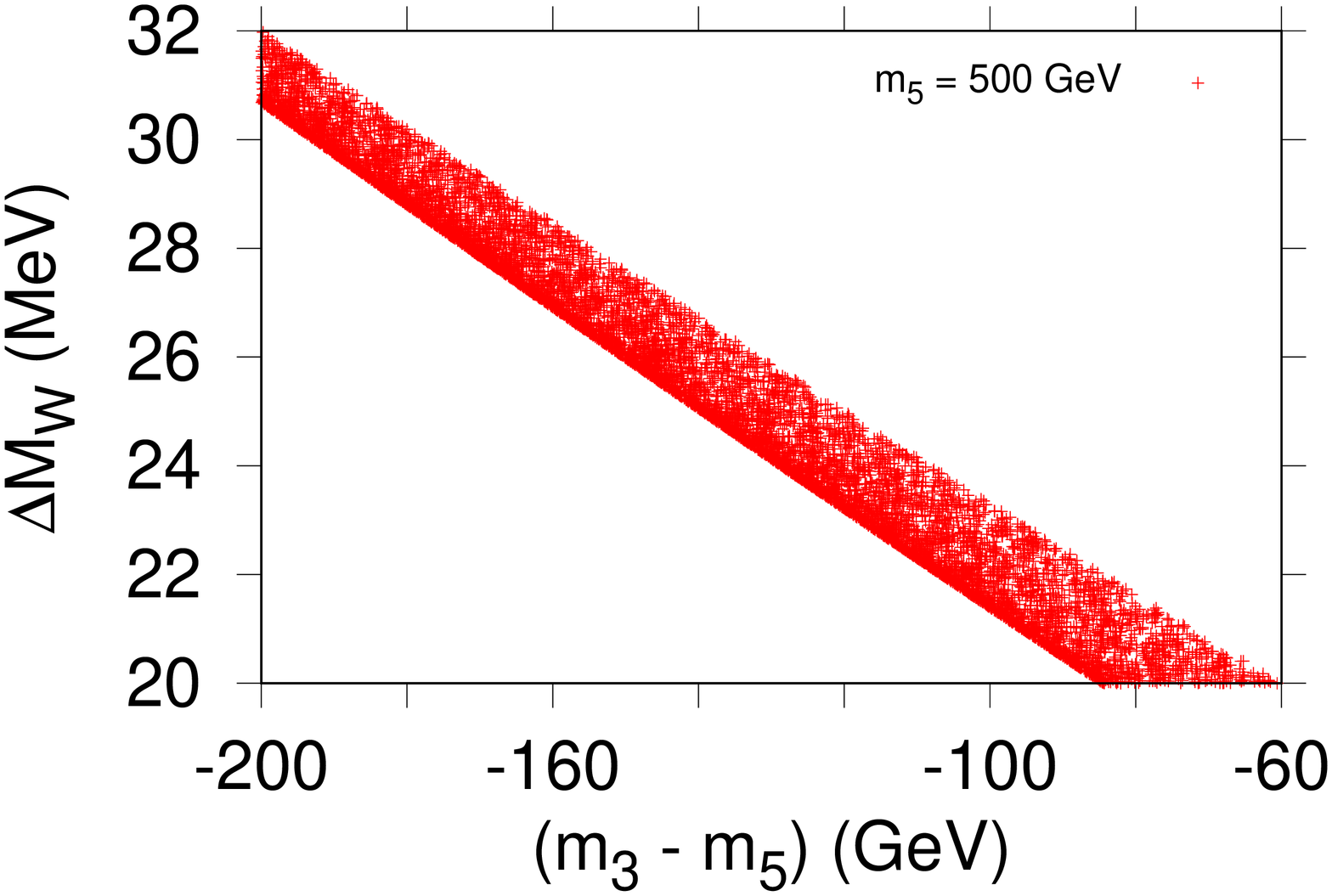} \\
\includegraphics[width=6cm, angle=270]{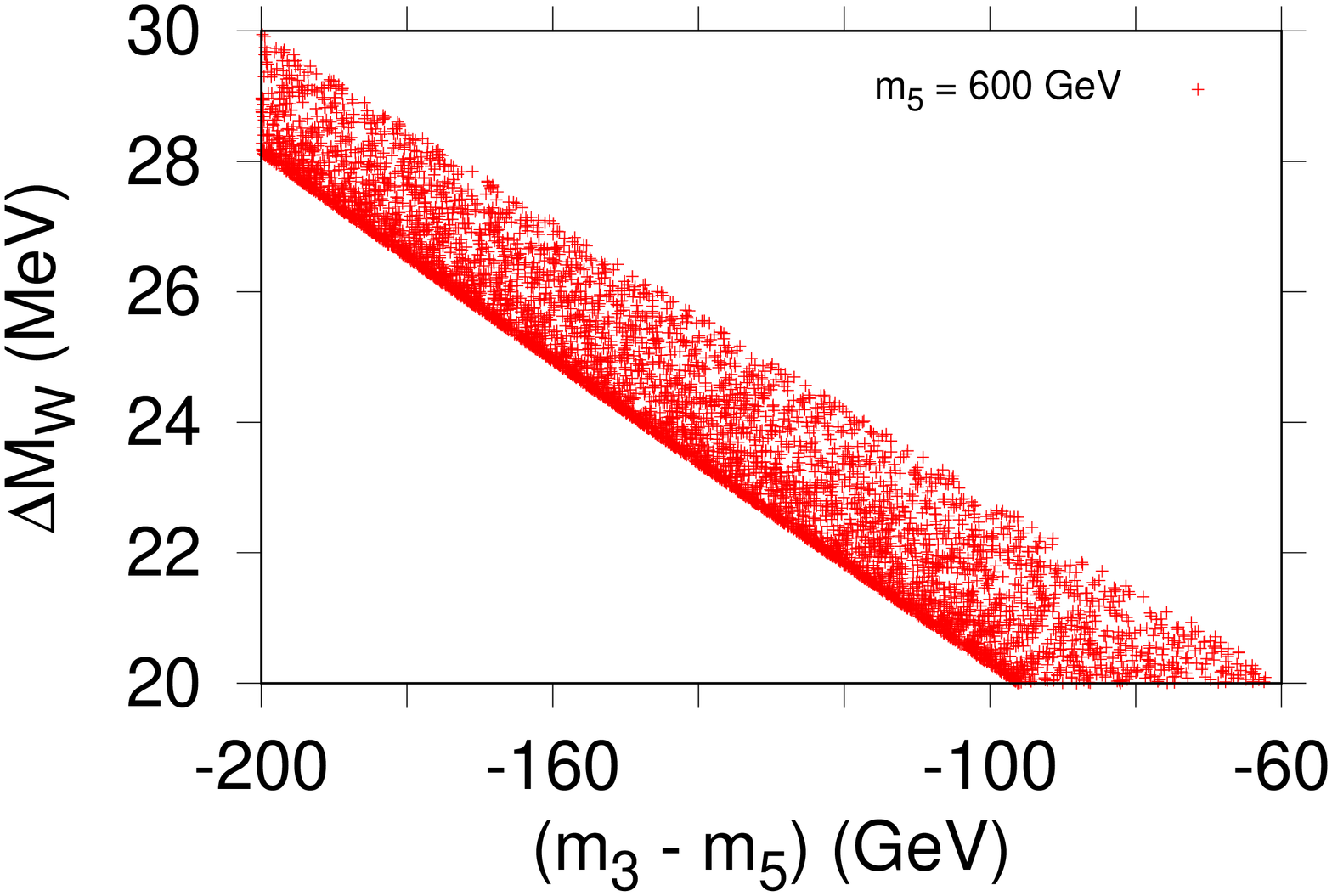} 
\includegraphics[width=6cm, angle=270]{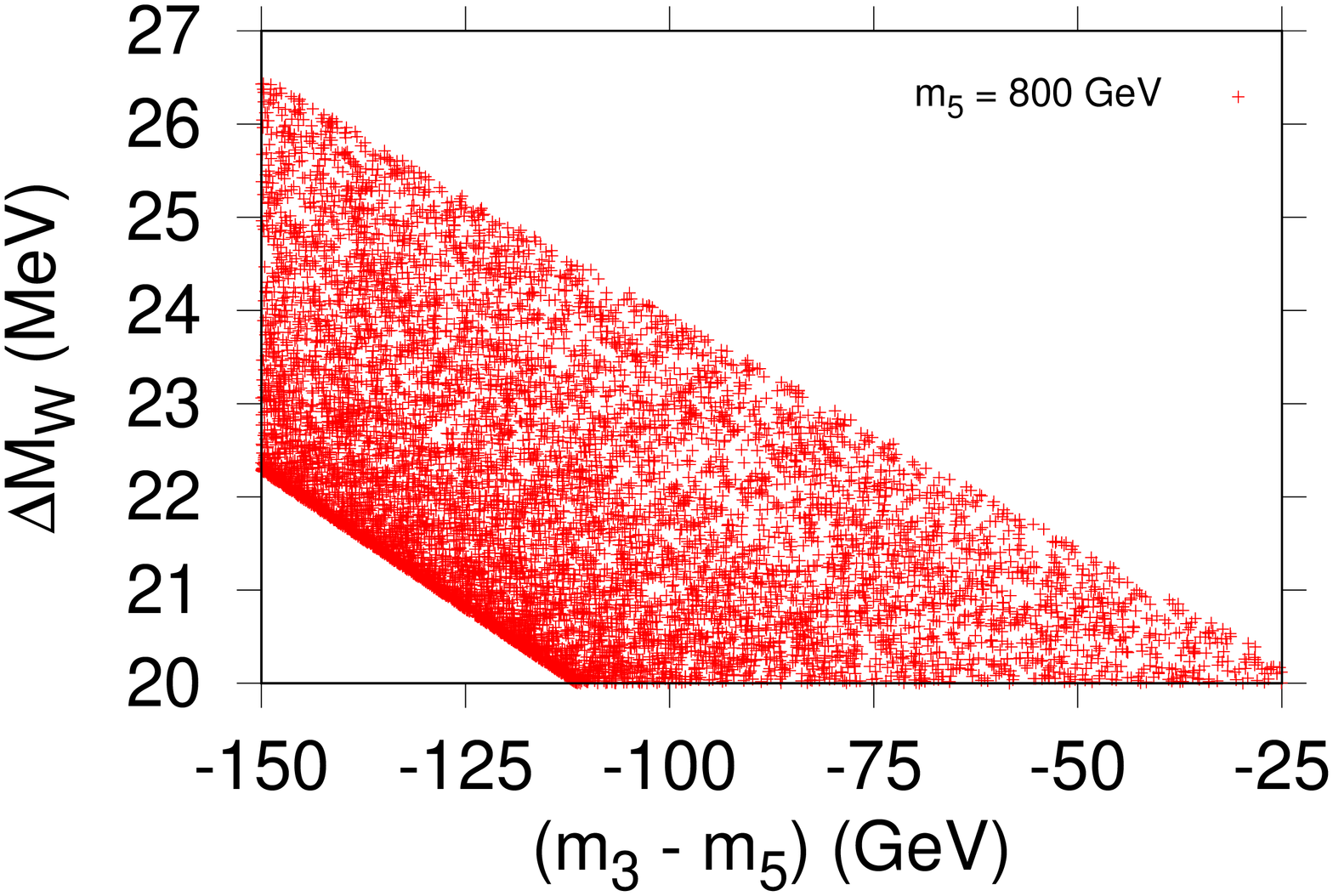} 
\caption{\small {Top left: Variation of $S$ with $m_3-m_5$ for $m_5=500$ GeV. Top right: Variation of 
$\Delta M_W$ with $m_3-m_5$ for $m_5 = 500$ GeV. Bottom left (right): The same as the top right plot,
but for $m_5=600 (800)$ GeV. In all these cases, the mixing angle $\vartheta$ as defined in Eq.\ (\ref{alpha}) 
has been kept fixed at $\sin\vartheta = 0.31$. The allowed region is generated by varying $v_\chi$.}}
\label{plot1} 
\end{figure}

Now let us see what happens if we lift the degeneracy between $m_3$ and $m_5$. The results are shown in 
Fig.\ \ref{plot1}, where we display the range of $\Delta M_W$ for various values of $m_3-m_5$, taking three
typical benchmark values for $m_5$, {\em viz.}, 500, 600, and 800 GeV. The band is generated by varying $v_\chi$, 
whose upper limit is again constrained from the di-gauge decay of $H^{++}$ \cite{CMS:2017fhs}. The maximum splitting 
between $m_3$ and $m_5$ is constrained by flavour observables  \cite{Hartling:2014aga}. The mass of the 
second custodial singlet $H$ has again been kept fixed at 150 GeV. 
To give the readers an idea of
how $S$ behaves, we have also shown its variation with $m_3-m_5$ for $m_5=500$ GeV. The nature is similar for 
other values of $m_5$. 

Thus, we find that in the GM model, $\Delta M_W$ may be enhanced by at most 32 MeV. This is within $1.6\sigma$ 
of the global average of $M_W$, and within $1\sigma$ if we neglect the $m_T(\mu)$ number from CDF. While this
is a reasonable agreement, it may not be enough if the GM parameter space gets constrained further, or if other
experiments confirm the CDF enhancement. As explicit violation of custodial symmetry (and thus introducing a 
nonzero $T$) goes against the spirit of the GM model, another viable alternative may be the extended GM model,
proposed in Ref.\ \cite{Kundu:2021pcg}. This model keeps $T=0$ at the tree-level, but new contributions to $S$ 
may emerge as the degeneracy among the custodial multiplets is lifted. This, however, makes the analysis
more complicated, and is beyond the scope of the present paper. 

%%%%%%%%%%%%%%%%%%%%%%%%%%%%%%%
\section{Conclusion}   \label{sec:conclude}
%%%%%%%%%%%%%%%%%%%%%%%%%%%%%%%

The recent CDF-II results on $M_W$ has a strong tension with the SM and, even when averaged with other
measurements of $M_W$, warrants new physics. Such new physics can be parametrised by the oblique parameters
$S$ and $T$, and the parameter space of any new physics model gets constrained by that. 

To keep the analysis simple, we focus on a particular class of model, namely, the Georgi-Machacek model, 
that keeps $T=0$ and hence $\Delta M_W$ depends only on $S$. 
The GM model with a heavy scalar spectrum is 
advantageous in the sense that it keeps $T=0$ by construction and $U\approx 0$ because of the heavy scalars, 
so the entire shift can be parametrised by the oblique $S$ parameter. 
We find that while the over-simplified version of GM (with $m_3=m_5$) fares rather poorly in explaining the tension, 
lifting the degeneracy between the custodial triplet and 5-plet helps in finding a reasonable agreement.

\section{Acknowledgements}

The author thanks Anirban Kundu for helpful discussion. 

%%%%%%%%%%%%%%% 

\end{document}